# Frequency comb vernier spectroscopy in the near infrared

F. Zhu,[1,*] J. Bounds,[1] A. Bicer,[1] J. Strohaber,[1] A. A. Kolomenskii,[1]

C. Gohle,[2] M. Amani,[3] and H. A. Schuessler[1,4]

[1] *Department of Physics and Astronomy, Texas A&M University, College Station, TX 77843-4242, USA*

[2] *Max-Planck-Institut für Quantenoptik, 85748 Garching, Germany*

[3] *Petroleum Engineering Program, Texas A&M University at Qatar, Doha 23874, Qatar*

[4] *Science Department, Texas A&M University at Qatar, Doha 23874, Qatar*

[*] *Corresponding author: zhuf@physics.tamu.edu*

We perform femtosecond frequency comb vernier spectroscopy in the near infrared with a femtosecond Er doped fiber laser, a scanning high-finesse cavity and an InGaAs camera. By utilizing the properties of a frequency comb and a scanning high-finesse cavity such spectroscopy provides broad spectral bandwidth, high spectral resolution, and high detection sensitivity on a short time scale. We achieved an absorption sensitivity of $\sim 8\times 10^{-8}$ cm$^{-1}$Hz$^{-1/2}$ corresponding to a detection limit of ~70 ppbv for acetylene, with a resolution of ~1.1 GHz in single images taken in 0.5 seconds and covering a frequency range of ~5 THz. These measurements have broad applications for sensing other greenhouse gases in this fingerprint near IR region with a simple apparatus.

## I. INTRODUCTION

Femtosecond frequency combs provide a regular comb structure of millions of narrow linewidth laser modes that can be accessed simultaneously [1]. In the past decade, a variety of methods were implemented to utilize this broad spectral bandwidth and high resolution comb structure for spectroscopy ranging from the extreme ultra violet to the terahertz region [2-10]. Among these, when combined with a traditional grating based monochromator, both the virtually imaged phased array (VIPA) spectrometer and the frequency comb vernier spectrometer were demonstrated with a charge-coupled device (CCD) camera in the wavelength range ~0.8 μm of a Ti:sapphire laser [5, 6]. Both spectrometers demonstrated sufficient resolution to resolve single comb modes with GHz comb spacing, and short data acquisition time to record two-dimensional images of a broad spectral range. With the ongoing development of frequency comb techniques, the VIPA spectrometer was extended to the Er:fiber laser wavelength range (~1.6μm) with an InGaAs camera [11-13] or a supercontinuum generated by a highly nonlinear fiber with an InSb camera [14], and to the mid-infrared range (3~4μm) via the optical parametric process and an InSb camera [15] for rapid detection of various gas species. To fulfill the high demands for sensitive and simultaneous detection of various trace gas species in many areas of science and technology, VIPA spectrometers were demonstrated to achieve high sensitivity for breath analysis [12] and trace impurities detection in semiconductor gases [14] by the steady state cavity enhancement techniques of locking frequency combs to high-finesse cavities to greatly increase the sample absorption length. However, the steady state cavity enhancement demands sophisticated optics and electronics to lock combs to high finesse cavities through servo systems, and even more importantly, the intracavity dispersion due to the highly reflective mirrors and the sample gases limits the transmitted bandwidth, in which the comb frequencies and cavity modes overlap. The frequency comb vernier spectroscopy is free of such limitations. By utilizing the interaction between a frequency comb and a *scanning* high-finesse cavity it provides broad spectral bandwidth, high spectral resolution, precise frequency calibration, and high detection sensitivity in a short time. The high sensitivity is obtained by using a simple scanning high-finesse cavity, which greatly extends the interaction length with the sample of interest and facilitates the improvement of the sensitivity. Potential hertz-level resolution can be reached, if the frequency comb is stabilized and scanned appropriately. Because of the scanning cavity approach, the requirement to accurately control the intracavity dispersion is relaxed, and the whole bandwidth of the comb can be transmitted through the cavity. An additional advantage of the vernier spectroscopy method is that it does not require the frequency comb to be locked to a high-finesse cavity, which makes it relatively simple and promising for gas analysis in field applications.

The principle of frequency comb vernier spectroscopy was first introduced in Ref. 6 and is briefly outlined below. The cavity length $L$ is intentionally adjusted to make the cavity free spectral range (FSR=$c/(2nL)$) mismatch the repetition rate $f_r$ of the frequency comb, so that only every $m$th comb mode is on resonance with every ($m$-1)th mode of the cavity, and the modes of the original frequency comb and the resonant frequencies of the cavity resemble a vernier with a ratio of FSR/$f_r$ =$m/(m$-1$)$. Figure 1 shows the case of FSR/$f_r$ =10/9. If the finesse of the cavity is high enough, the other off resonance comb modes will be strongly suppressed, which results in an effective spectral filter for the original comb. The frequency comb filtered by the cavity has a mode spacing of $mf_r$, thus for a large number $m$, it can be resolved with a simple grating based spectrograph. Access of all original comb lines can be achieved by scanning the cavity length $L$ to shift cavity modes, so that the next set of $mf_r$ spaced comb modes can be lined up with FSR resonances of cavity modes and transmitted through the cavity. After scanning $L$ by half of wavelength (i.e., one FSR), $m$ groups of $mf_r$ spaced combs have been selected and transmitted through the cavity, and the initial group starts again at a slightly changed vernier ratio.

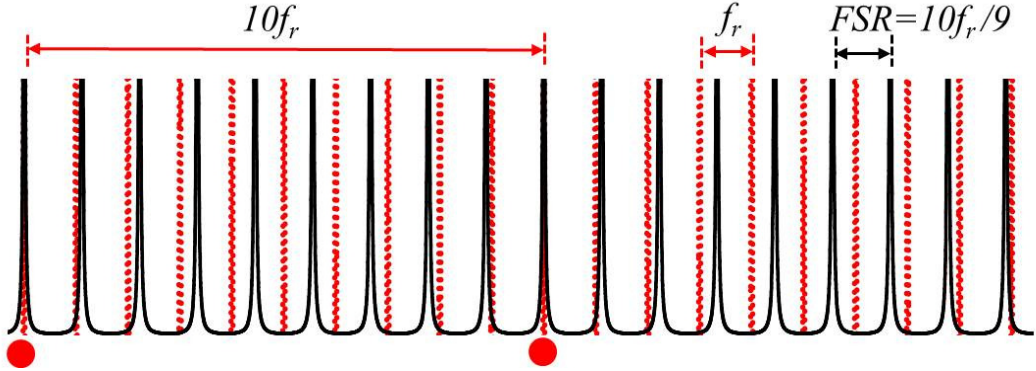

FIG. 1. (Color online) Scheme of frequency comb vernier spectroscopy for the vernier ratio of FSR/ $f_r$ =10/9. Red dotted lines are frequency comb modes, black solid lines are cavity resonant modes. Large red dots mark the transmitted comb modes.

A tilting mirror driven by a Galvo motor, which was synchronized with the scanning cavity, was used to map the different $mf_r$ spaced groups onto one spatial dimension of the detector array, resolving the single frequency comb modes of 1 GHz spacing on a CCD camera [6]. The principle of the frequency comb vernier scheme was also demonstrated between a passively mode-locked Er:fiber laser and a fiber ring resonator with an optical fiber strain sensor [16]. In addition, a vernier scheme between signal and idler was employed in the nested cavities of an optical parametric oscillator to tune the wavelength of the laser source for spectroscopic applications [17].

In this work, we extend the femtosecond frequency comb vernier spectroscopy to the Er:fiber laser wavelength range in the near infrared. With a higher finesse cavity, the achieved sensitivity increases almost by two orders of magnitude compared to the original demonstration [6].

## II. EXPERIMENTAL SETUP

The experimental setup is presented in Fig. 2.

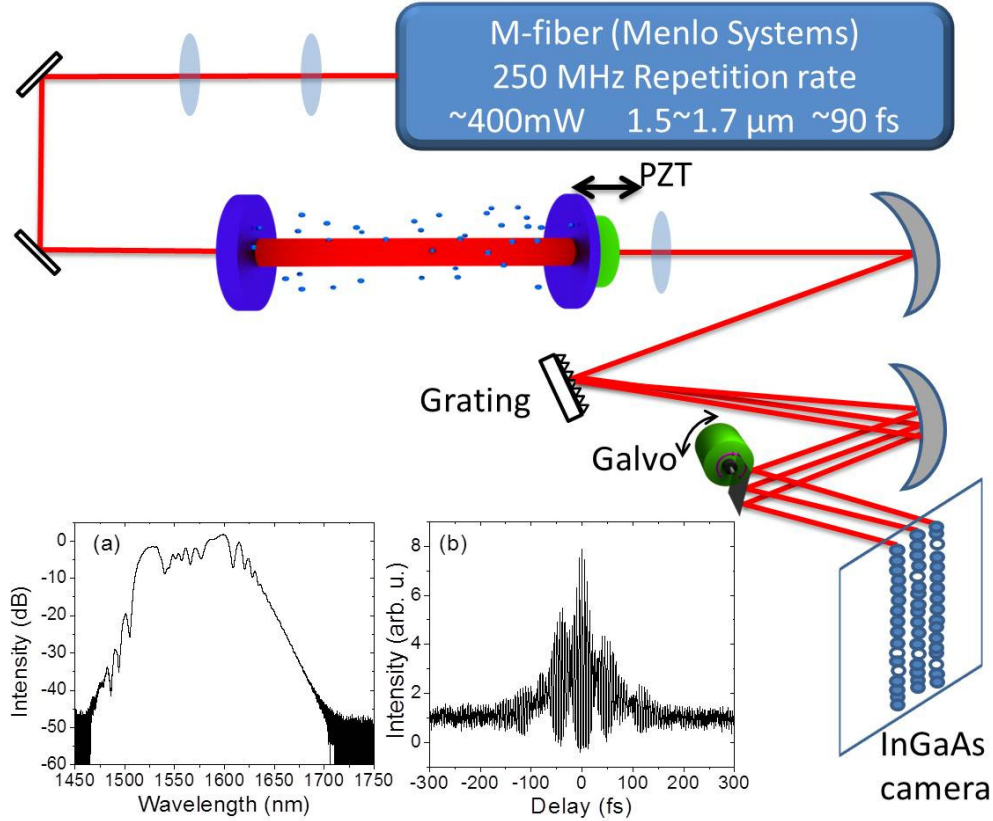

FIG. 2. (Color online) The experimental setup includes a near infrared frequency comb source, mirrors and mode matching lenses, a scanning high-finesse cavity, a home-built Czerny-Turner spectrometer, a tilting mirror driven by a Galvo, and an InGaAs camera. Insets: (a) spectrum and (b) interferometric autocorrelation trace of the femtosecond Er:fiber laser.

We employ a femtosecond Er:fiber laser (Menlo Systems GmbH, M-fiber) as a frequency comb source and an InGaAs camera (Princeton Instruments, PIoNIR 640) for detection. The Er:fiber laser consists of an oscillator and an amplifier. Amplified pulses are coupled into a highly nonlinear fiber to generate a frequency comb ranging from 1500 nm to 1700 nm. The total output power is ~400 mW. The pulse duration is ~90 fs. The repetition rate $f_r$ is stabilized at 250 MHz, and the carrier-envelope offset frequency $f_o$ is free-running. After mode matching lenses, the frequency comb is coupled into a scanning high-finesse Fabry-Perot cavity consisting of a flat mirror and a 2 m radius concave mirror. The flat mirror is mounted on a high precision translational stage (Newport, GTS150) to accurately control the cavity length $L$ to impose different vernier ratios; the concave mirror is mounted on a piezoelectric transducer (PZT) tube to scan $L$ more than half of one wavelength (i.e. one FSR). Both cavity mirrors (LAYERTEC GmbH) have the high reflectivity of R~99.99%, corresponding to a high-finesse of ~30000, which we confirmed with a traditional continuous wave (CW) cavity ring down measurement. We use a 5×5 cm$^2$, 300 grooves per millimeter gold coated grating (McPherson) in a homebuilt 30 cm focal length Czerny-Turner spectrometer. A tilting mirror driven by a Galvo, which is synchronized with the scanning cavity, is used to map different groups of filtered combs onto one spatial dimension of the InGaAs camera. The PIoNIR 640 (now NIRvana 640) camera from Princeton Instruments features a two-dimensional 640×512 InGaAs focal plane array. Pixel size is 20×20 µm$^2$. The camera exposure time is also synchronized with the PZT scanning and Galvo tilting to record one single scan image. The detector array is thermoelectrically cooled to $-80^0$C to reduce the dark current. The red end cut-off wavelength of the InGaAs camera shifts towards the blue by ~ 8 nm per every 10$^0$C temperature drop. Experimentally we found that the red end cut-off wavelength at $-80^0$C is around 1600 nm, and that a signal above 1550 nm is weak as the quantum efficiency drops between 1550 nm and 1600 nm. We chose acetylene as a sample, since it has an absorption band v1+v3 between 1510 nm and 1550 nm), which conveniently lies within the sensitivity range of the camera.

## III.  RESULTS

In the experiments, we first fill the cavity with a mixture of ~5 ppmv acetylene and air at room temperature and atmospheric pressure to take the sample image. Then we use compressed air to purge the cavity and take the reference image under the same conditions. The images with 500 ms scan time have good intensity contrast. By tuning the cavity length *L*, the vernier ratio could be changed; typical reference and sample images with two different vernier ratios of ~500/499 and ~250/249 are presented in Fig. 3.

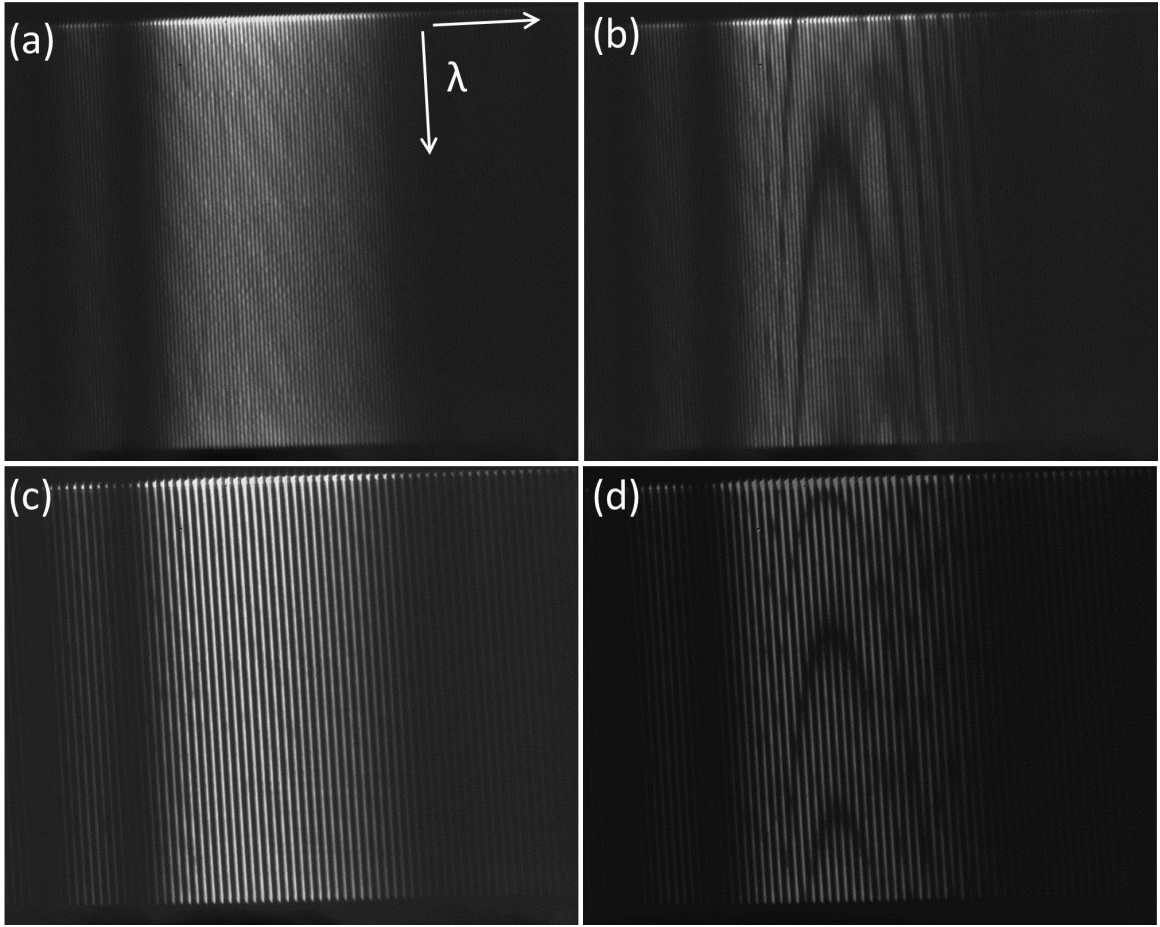

FIG. 3. Vernier reference and sample images: (a) only air, (b) 5 ppmv acetylene, vernier ratio ~250/249; (c) only air, (d) 5 ppmv acetylene, vernier ratio ~500/499. The varying brightness along the *x* axis is caused by the non-uniform spectral density distribution of the near infrared frequency comb source; for the *y* axis, especially at the top of images (beginning of the scan), it is caused by the nonlinearity of the PZT and Galvo scans. Along the *y* axis, the acetylene absorption features are quite clear for both vernier ratios.

When the vernier ratio is changed from ~500/499 to ~250/249 by tuning the cavity length *L*, the mode spacing of the filtered frequency comb is decreased from $500f_r$ to $250f_r$, corresponding to 125 GHz and 62.5 GHz respectively, and the number of observed fringes in images doubles.

The resolution of the grating based Czerny-Turner spectrometer is experimentally determined to be around 40 GHz by further reducing the mode spacing of the filtered frequency comb to $150f_r$ (vernier ratio ~150/149), in which case the filtered modes of 37.5 GHz spacing were not resolved. Only images with mode spacing of the filtered comb above the resolution of Czerny-Turner spectrometer show distinct fringes. The spot size imaged on the camera is ~50 µm, so that each resolved frequency component is sampled by 3×3 pixels. Since there are about ~340 vertical pixels in one FSR scan, and for the vernier ratio of ~500/499, the filtered comb spacing is 125 GHz, and the resolution under these conditions is 125 GHz/[(340 pixels)/(3 pixels)] ~1.1 GHz, corresponding to about 4 to 5 comb modes. For the vernier ratio of ~250/249, the resolution is ~550 MHz, corresponding to about 2 to 3 comb modes. Thus, for the resolved frequency components of close frequencies, the intensities are higher for the vernier ratio of ~500/499 than for those in the ~250/249 case. This can be seen in Fig. 3, where the intensities of the fringes are brighter, and the image contrasts between fringes and backgrounds are better in Fig. 3 (c) and (d) than in Fig. 3 (a) and (b). Even though the single comb mode of 250 MHz mode spacing is not resolved under current experimental conditions, the acetylene absorption features are quite clear for both vernier ratios, since the widths of absorption lines are several GHz.

We process the images with a procedure similar to Ref. 12, as is depicted in Figure 4. A pair of reference and sample images with the same vernier ratio is used to retrieve the normalized spectral absorption information. In the first step, the locations of fringe centers are identified with the reference image. Then we average a pixel matrix (at least 3×3) around fringe centers to generate the fringe intensity arrays for both reference and sample images. Next we normalize the sample fringe array to the reference fringe array to calculate the sample absorption by using $α=−1/(2L)·\ln[1/R^2·(1−(1−R^2)/(I_{sample}/I_{ref}))]$ [14]. By comparing two images, an absorption image is generated. Since in frequency comb vernier spectroscopy, after scanning one FSR, the initial group starts to repeat with a one-step shifted fringe position, a unique data set can be identified from the absorption image. After determining the vertical boundaries of the unique data set, the two-dimensional image is unwrapped into a traditional one-dimensional absorption spectrum.

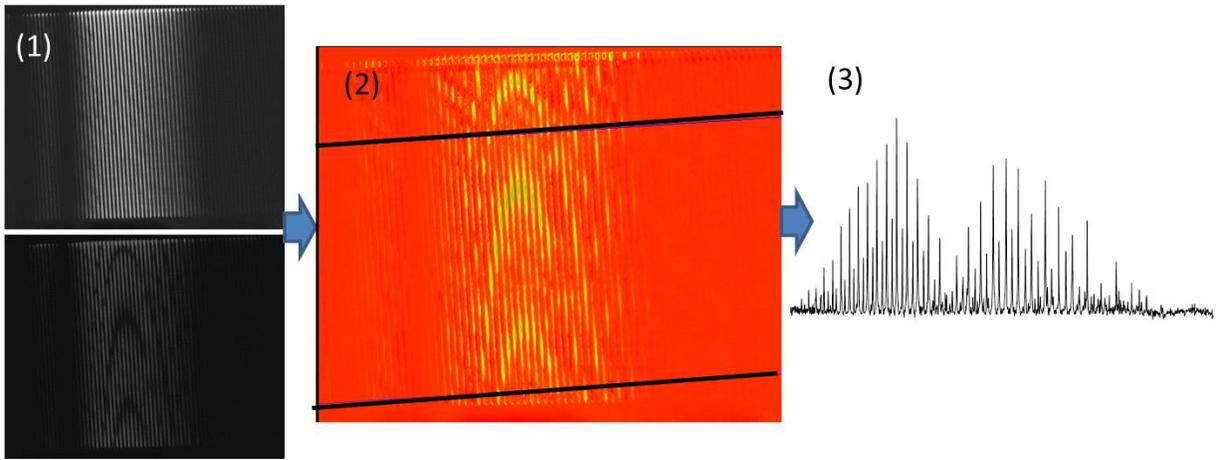

FIG. 4. (Color online) Image processing procedure for generating an absorption spectrum from a pair of reference and sample images: a pair of reference and sample images (1) is used to generate an absorption image (2); after determining the vertical boundaries for a unique data set (two black lines across the absorption image), the fringes between the boundaries are unwrapped into a traditional one dimensional absorption spectrum (3).

With the information about the vernier ratio and the mode spacing of the filtered frequency comb, we use one prominent absorption peak of acetylene from the HITRAN database [18] to calibrate the spectrum. The calibration accuracy is limited by the resolution of the vernier spectrometer, assuming a linear scan of PZT and Galvo in the unique data set portion of the image. Since under current experimental conditions, single comb modes of 250 MHz spacing are not identified, the exact frequency calibration of the comb lines is not carried out, and consequently, the generated spectra are distorted by the intracavity dispersion. Experimentally, the spectroscopic calibration can be carried out with a narrow-linewidth CW laser with known wavelength by spatially overlapping with the frequency comb and coupling them together into the cavity. Although the ~250/249 vernier ratio indicates a better spectral resolution of ~550 MHz, the intensity of each resolved frequency element is lower than that of the ~500/499 vernier ratio, and the lower intensity of fewer frequency comb modes covered in one resolved frequency component has more fluctuations. Hence, the root mean square noise level at positions with no absorption appears higher than that for ~500/499 vernier ratio. The absorption signals of both vernier ratios are nearly the same, as expected. Thus, the ~500/499 vernier ratio images render a better signal-to-noise ratio (SNR~80 compared to the SNR~15 in ~250/249 vernier ratio case for the strongest absorption line). A comparison between the normalized spectrum retrieved from the images with ~500/499 and ~250/249 vernier ratio and the HITRAN simulation is presented in Fig. 5. The retrieved spectrum shows good agreement with the literature values in frequency scale and amplitude, and it covers the spectral range from 1510 to 1550 nm, i. e. about 5 THz. The spectral range is largely limited by the dimension and sensitivity of the InGaAs camera detection array under current experimental conditions.

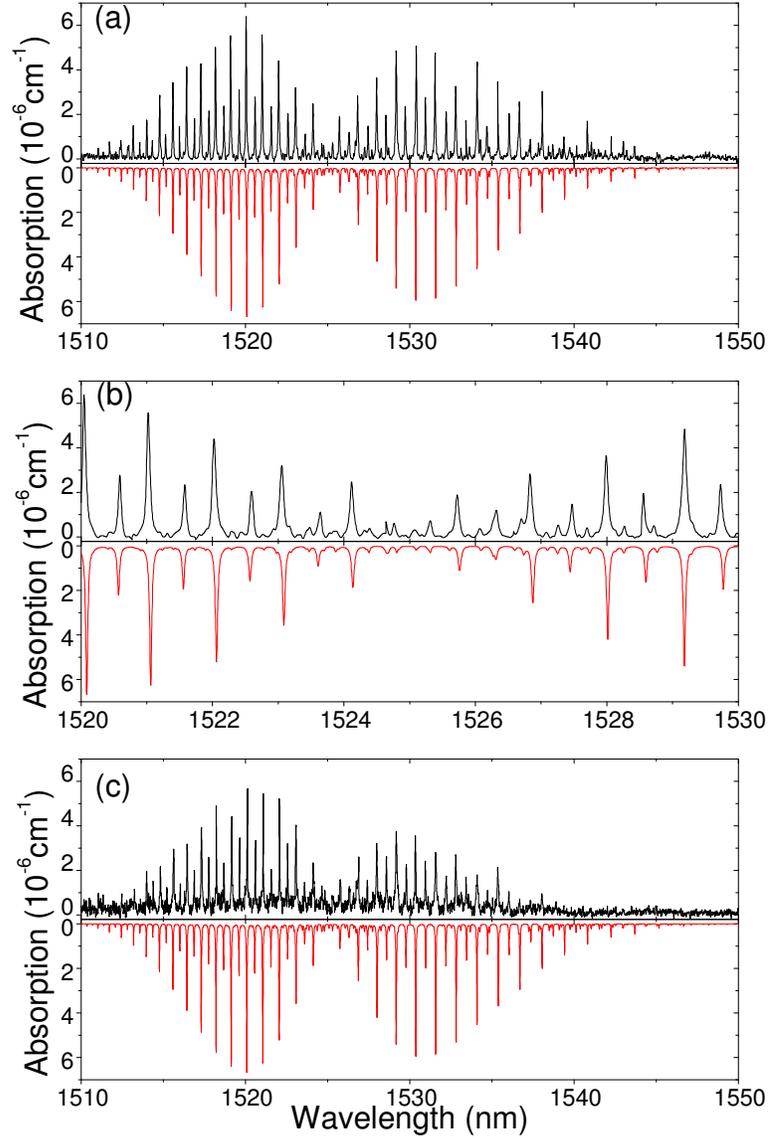

FIG. 5. (Color online) Acetylene absorption spectrum retrieved from images of ~500/499 and ~250/249 vernier ratio (black) and comparison with HITRAN simulation of 5 ppmv acetylene at room temperature and atmospheric pressure (red, inverted for clarity): (a) in the broad range between 1510~1550 nm and (b) expanded view between 1520~1530 nm for a vernier ratio ~500/499; (c) in the broad range between 1510~1550 nm for a vernier ratio ~250/249. Due to the reduced power of the resolved frequency elements the signal-to-noise ratio in this case is lower compared to the case of ~500/499 vernier ratio.

## IV.  DISCUSSION AND SUMMARY

By evaluating the root mean square noise level at positions where no absorption signal is detected, we estimate the absorption sensitivity at the current experimental conditions to be $\alpha_{min}$~8×10$^{-8}$ cm$^{-1}$ for the vernier ratio of ~500/499. Considering the strongest absorption line of acetylene, this corresponds to a detection limit about ~70 ppbv of acetylene concentration under room temperature and atmospheric pressure conditions. The noise equivalent absorption (NEA) is defined as $\alpha_{min}(T/M)^{1/2}$, where the total acquisition time $T$ is 1 s, including both the sample and reference images, and $M$ is the number of resolved spectral elements, estimated as the total bandwidth of the measurement divided by the spectral resolution, i.e. 5 THz /1.1 GHz~ 4500. Consequently, the NEA is ~ 1.2×10$^{-9}$ cm$^{-1}$Hz$^{-1/2}$ per spectral element for the current experimental conditions. This sensitivity level is about two orders of magnitude above the shot noise limit [7, 8] and is due to the mechanical noise, laser intensity noise, etc. The mechanical noise, such as acoustic vibrations, disturbs the high-finesse cavity, which causes a slight nonlinearity of the cavity scan and results in changes of the integration times, intensities and positions of recorded fringes of spectra. Also because of the pulse to pulse jitter, the intensity profile fluctuates, which adds noise to different portions of the normalized spectrum, since the absorption and reference spectra are not recorded simultaneously. Provided an acoustic noise isolated cavity, a fully stabilized frequency comb and good linear control over the scanning elements of PZT and Galvo, it would be possible to average several vernier images to further reduce noise and increase the detection sensitivity.

In summary, we demonstrate femtosecond frequency comb vernier spectroscopy in the Er:fiber laser wavelength range with an InGaAs camera. These measurements have broad applications for sensing greenhouse gases in this fingerprint near IR region with a simple apparatus. We achieved an absorption sensitivity of ~8×10$^{-8}$ cm$^{-1}$, corresponding to a detection limit of ~70 ppbv for acetylene, with a resolution of ~1.1 GHz by using single images taken in 0.5 s and covering a frequency range of ~5 THz. Provided a higher resolution spectrometer with a larger grating and a spot size better matching the camera pixel size, it seems possible to resolve the single frequency comb modes of a few hundreds MHz spacing with the future development of cameras. An image with resolved single comb modes not only provides precise calibration of the frequencies, but also can be fitted to retrieve phase information [6].

AKNOWLEDGMENTS

We thank Princeton Instruments for the generous loan of the PIoNIR 640 camera and the Robert A. Welch Foundation for funding provided under grant No. A1546. This publication was made possible by the NPRP award [NPRP 5 - 994 - 1 - 172] from the Qatar National Research Fund (a member of The Qatar Foundation). The statements made herein are solely the responsibility of the authors.